\documentclass[12pt]{article}
\usepackage{amsmath}
\usepackage{graphicx}
\usepackage{enumerate}
\usepackage{amsmath,amssymb}               
\usepackage{amsthm}
\usepackage{graphics}
\usepackage{soul,color}
\usepackage{xcolor}
\usepackage{subcaption}
\usepackage{multirow}

\newtheorem{proposition}{Proposition}

\newtheorem{example}{Example}
\usepackage{DNstyle}

\newcommand\indep{\protect\mathpalette{\protect\independenT}{\perp}}
\def\independenT#1#2{\mathrel{\rlap{$#1#2$}\mkern2mu{#1#2}}}

\usepackage[normalem]{ulem}
\usepackage{natbib}

\usepackage{url} 

\newcommand{\blind}{1}

\begin{document}

	\def\spacingset#1{\renewcommand{\baselinestretch}%
		{#1}\small\normalsize} \spacingset{1}

	
	\if1\blind
	{
		\title{\bf Causal inference for semi-competing risks data}
		\author{Daniel Nevo and 	Malka Gorfine\thanks{
\textit{The authors gratefully acknowledge  the Adults Changes in Thought Study (ACT) team for sharing the data for the motivating example. The ACT was funded by the National Institute on Aging (grant number: U01AG006781). MG gratefully acknowledge support from the U.S.-Israel Binational Science Foundation (BSF, 2016126), and the Israel Science Foundation (ISF, 1067/17) in carrying out this	work.}}\hspace{.2cm}\\
			Department of Statistics and Operations Research, Tel Aviv University}
		\maketitle
	} \fi
	
	\if0\blind
	{
		\bigskip
		\bigskip
		\bigskip
		\begin{center}
			{\LARGE\bf Causal inference for semi-competing risks data}
		\end{center}
		\medskip
	} \fi
	
	\bigskip
	\begin{abstract}
An emerging challenge for time-to-event data is studying semi-competing risks, namely when two event times are of interest: a non-terminal event time (e.g. age at disease diagnosis), and a terminal event time (e.g. age at death). The non-terminal event is observed only if it precedes the terminal event, which may occur before or after the non-terminal event. Studying treatment or intervention effects on the dual event times is complicated because for some units, the non-terminal event may occur  under one treatment value but not under the other. Until recently, existing approaches (e.g., the survivor average causal effect) generally disregarded the time-to-event nature of both outcomes. More recent research focused on principal strata effects within time-varying populations under	 Bayesian approaches. 
In this paper, we propose  alternative  non time-varying estimands, based on a single stratification of the population.   We  present a novel assumption utilizing the time-to-event nature of the data, which is weaker than the often-invoked monotonicity assumption.  We derive results on partial identifiability,   suggest a sensitivity analysis approach, and give conditions under which full identification is possible. Finally, we present non-parametric and semi-parametric estimation methods for right-censored data. 
	\end{abstract}
	
	\noindent%
	\vfill
	
	\newpage
	\spacingset{1.5} 

\section{Introduction}
 \label{Sec:Intro}
Semi-competing risks data  arise when  times to two events of interest are studied,  non-terminal (e.g. disease diagnosis) and  terminal  (e.g. death) events. The non-terminal event  is observed only if it precedes the terminal event, which may occur before or after the non-terminal event.

When the goal is to assess the causal effect of  a treatment, exposure or an  intervention on the two events,  the effect on the terminal event only might not capture the scientific effect of interest. As a motivating example,   consider the complex effect of having at least one Apolipoprotein E $\epsilon4$ allele (APOE) on late-onset Alzheimer's disease (AD) diagnosis (non-terminal event) and death (terminal event). APOE is a well-established risk factor for AD \citep{alzheimer20192019}. However, it is unknown whether APOE has a distinct causal effect on survival \citep{corder1993gene,dal2002apoe,wang2015genetic}, or even how to define this effect, due to its entanglement with  AD, that by itself affects life expectancy. An observed APOE effect on death  might be  the result of a strong APOE effect on AD, and the occurrence of AD  expedited the  terminal event occurrence. An alternative explanation is that APOE affected death directly.  

Problems  also arise for the  causal effect of APOE on AD (the non-terminal event). The standard average treatment effect is ill-defined because some study participants die  without an AD diagnosis, leading to an infinite AD event time.  Motivated by these challenges,  this paper  studies causal effects for semi-competing risks data, including their definition, non-parametric (partial) identification, estimation and inference about such effects.




Recent advancements in semi-competing risks data analysis often consider one of two main approaches, copula-based methods and the illness-death model.   The copula-based approach focuses on the identifiable region of the  joint survival function of the event times, namely the region in which the terminal event occurs after the non-terminal event \citep{fine2001semi}. Then, a parametric copula function of the two marginal distributions is used to model the joint survival function on this region.  While regression methods  were also developed under this framework \citep{peng2007regression,hsieh2008regression},  to our knowledge, causal  interpretation of this approach has not been studied. The illness-death model approach  \citep{andersen1993statistical} focuses on  transition times between three different states:  An initial state  (state ``0'', healthy), a diseased state (``1'', disease), and an absorbing state (``2'', death). Possible transitions are $0\rightarrow1$, $0\rightarrow2$ and $1\rightarrow2$.   Various specifications and time-to-event models have been suggested to analyze semi-competing risks data in this framework \citep{meira2006nonparametric,xu2010statistical}.  Typically, inter-personal dependence between illness and death times is  assumed to be captured by covariates, or modeled via a random effect, known as frailty \citep{xu2010statistical,lee2017accelerated,gorfine2020marginalized}.   

In the causal inference paradigm, the related  ``truncation by death'' problem has been studied  \citep{zhang2003estimation,hayden2005estimator, zhang2009likelihood,long2013sharpening,tchetgen2014identification,ding2017principal}. This problem arises when researchers aspire to compare  outcomes (e.g. quality of life index) between the treated and untreated groups at a fixed time after the treatment was assigned, but  some study participants have died prior to the outcome collection, hence the average causal effect is not well-defined. Using  principal stratification \citep{frangakis2002principal}, the survivor average causal effect (SACE) is the causal effect within the stratum of people who would have survived under both treatment values.  A similar problem arises in vaccination studies, where the effect of a vaccine on post-infection outcomes is  well-defined only among those who would have been infected with or without vaccinating \citep{shepherd2007sensitivity}.  Generally, the SACE is non-parametrically unidentifiable, unless strong assumptions are made \citep{hayden2005estimator,zhang2009likelihood}. Therefore, researchers have developed bounds \citep{zhang2003estimation,long2013sharpening} and sensitivity analyses
\citep{hayden2005estimator,shepherd2007sensitivity}. However, none of these papers apply to  semi-competing risks data,  where the time-to-event nature of  both disease diagnosis and survival times must be addressed.  

\cite{robins1995analytic} utilized additional assumptions to show the average causal effect (not SACE) is identifiable when
censoring time is always observed, and without distinguishing between censoring due to death or other
reasons. More recently, two papers have considered causal questions for semi-competing risks data \citep{comment2019survivor,xu2020bayesian}. These papers suggested Bayesian methods to estimate the  time-varying SACE (TV-SACE)  and time-varying restricted mean survival time (RM-SACE).  While these papers provide major improvement towards causal reasoning for semi-competing risks data, their proposed estimands can be hard to interpret, because at each time $t$ the population for which the time-varying estimands are defined is changing. Additionally,  one of the works rely on parametric frailty-based illness-death models \citep{comment2019survivor}, while the other coupled a parametric sensitivity copula-based analysis with  a non-parametric Bayesian method under an additional identifying assumption \citep{xu2020bayesian}. Questions of  non-parametric partial identifiability, including non-parametric bounds under weaker assumptions, or sensitivity analyses without the copula model, as well  as frequentist  non-parametric and semi-parametric  estimation  were all not addressed.  

In this paper, we define new causal estimands for semi-competing risks data. The general idea resembles principal stratification in that we propose to stratify the population according to their potential outcomes. However, unlike the TV-SACE and RM-SACE, we focus on effects defined within a fixed, not time-varying, population. The contributions of our paper can be summarized as follows:
\begin{itemize}
	\item New causal estimands for semi-competing risks data.
	\item Novel assumptions suitable for the semi-competing risks data structure. For example, our new assumption termed \textit{order preservation} is weaker than the traditionally-used monotonicity assumption \citep{zhang2003estimation}. 
	\item Partial and full identifiability results for the proposed estimands.
	\item New tools for casual inference from semi-competing risks data that can be used in the presence of right censoring, a ubiquitous challenge in the analysis of time-to-event data. These include bounds, sensitivity analyses,  non-parametric and semi-parametric estimators and inference methods.
	\item The \textbf{R} package \texttt{CausalSemiComp}  implementing the developed methodology. Simulation results are fully reproducible. Both the package and reproducibility materials are available from the first author's Github account.
\end{itemize}


 

\section{Notation and causal estimands}
\label{Sec:NotationEstimands}

Let $T_1(a)$ and $T_2(a)$ be the times to disease diagnosis and  death, respectively,  under treatment level $A=a$. Throughout the paper, we call $A$ a treatment, although in practice it could also be an exposure or an intervention.  For example,  $A=1$ may indicate having at least one APOE $\epsilon4$ allele, while $A=0$ indicates no APOE $\epsilon$4 alleles.  Then, $T_1(a)$ and $T_2(a)$ are the potential Alzheimer's diagnosis and death times had we intervened on the APOE gene and set it to $A=a$.   

The assumed probability space for $\{T_1(0),T_2(0), T_1(1), T_2(1)\}$ is as follows. Let $f_a(t_1,t_2)$ be the joint density function of $\{T_1(a),T_2(a)\}$ on $0<t_1\le t_2$ for $a=0,1$. Because the  terminal event may occur before the non-terminal event, precluding the observation of the latter, then $\Pr[T_1(a) <\infty] =\int_{0}^{\infty}\int_{0}^{t_2}f_a(t_1,t_2)dt_1dt_2<1$.  The event $T_1(a)=\infty$ or, stated differently, $T_1(a)>T_2(a)$ corresponds to not having the non-terminal event under  $A=a$.  We defer assumptions on the joint distribution of $\{T_1(0),T_2(0)\}$ and $\{T_1(1),T_2(1)\}$ for later; these two pairs could never be observed simultaneously.  

The causal inference perspective often focuses on contrasting averages of potential outcomes under different treatment levels. Consider the average causal effect of APOE on the survival time $E[T_2(1)-T_2(0)]$. Because APOE affects AD incidence  \citep{alzheimer20192019}, and because AD patients are likely to die earlier compared to the scenario they did not have AD \citep{tom2015characterization}, we would expect $E[T_2(1)-T_2(0)]< 0$, even if APOE does not carry an effect on age at death other than the effect resulting from earlier AD onset. Additionally, when focus is on the non-terminal event, $E[T_1(1)-T_1(0)]$ is  ill-defined, because $T_1(a)=\infty$ for  some study units for either or both $a=0$ and $a=1$.

Therefore, causal inference for semi-competing risks data is entangled. Principal stratification effects  suggested  to address this challenge are defined  within a strata of the population created by the potential values of a  post-randomization variable,   which is not the object of inquiry \citep{frangakis2002principal}. For semi-competing risks, principal stratification approaches can be divided into two types:  time-fixed and time-varying estimands. The time-fixed approach focuses on a single time point $\overline{t}$, say age 80, and define the binary potential outcomes $\mathcal{S}(a)=I\{T_2(a)>\overline{t}\}$ and $\mathcal{Y}(a)=I\{T_1(a)\le \overline{t}\}$ as the survival status and Alzheimer's status at time $\overline{t}$ under treatment level $A=a$. Then, the problem is posed as a truncation-by-death problem with the target causal estimand being the SACE: $E[\mathcal{Y}(1)-\mathcal{Y}(0)|\mathcal{S}(1)=\mathcal{S}(0)=1]$. The SACE compares the population-level  risk of AD by age $\overline{t}$ with and without APOE  only among those who would have survived until age $\overline{t}$ under both APOE statuses. Typically, strong assumptions are needed for identifying the SACE and researchers presented a variety of analyses under various assumptions. One commonly-made assumption is the \textit{monotonicity} assumption $\mathcal{S}(1)\le\mathcal{S}(0)$ \citep{zhang2003estimation,ding2011identifiability,ding2017principal}. That is, those surviving beyond time $\overline{t}$ and have APOE would also survive beyond $\overline{t}$ had they did not have APOE.   A more recent approach \citep{comment2019survivor,xu2020bayesian} focuses on the TV-SACE causal effect 
$$TV\text{-}SACE(r,t)=\Pr[T_1(1)<r|T_2(1)>t,T_2(0)>t] - \Pr[T_1(0)<r|T_2(1)>t,T_2(0)>t]$$
for $r\le t$. The $TV$-$SACE$ has two major advantages over the standard SACE: at each time $t$ it compares  non-terminal event rates in multiple time points, and it respects that the population of always-survivors depends on the chosen time $t$. However, exactly because the population changes with time, it can be hard to interpret it as a function of $t$. A change in  $TV$-$SACE(r,t)$ as a function of $t$ may reflect a change in the population or a change in efficacy of the treatment. Furthermore, that treatment effects cannot be summarized into a single number, even for a fixed $r$, might create a challenge in communicating the results to a broader audience. 

\subsection{Population stratification and alternative estimands}

We propose alternative estimands, based on a single partition of the  population.  We divide the population into four strata according to their illness and death ``order''. This partition  resembles the principal stratification framework that typically stratifies the population according to a post-treatment variable which is not the outcome. Therefore,  because our partition involves the dual outcomes,  we term it  \textit{population stratification}.  Define the following four subpopulations with respect to  the order of $T_1(a),T_2(a)$ for $a=0,1$. The ``always-diseased'' ($ad$)  will be diagnosed with the disease at some point of their life, regardless of their treatment status ($T_1(a)\le T_2(a)$ for $a=0,1$); the ``never-diseased'' ($nd$) will die without having the disease first, regardless of their treatment status ($T_1(a)>T_2(a)$ for $a=0,1$); the ``harmed'' ($dh$) would be diagnosed with the disease prior to death only if treated  ($T_1(0)>T_2(0),T_1(1)\le T_2(1)$); and the ``protected'' ($dp$) would be diagnosed with the disease only if untreated  ($T_1(0) \le T_2(0),T_1(1)>T_2(1)$).  Let $\pi_{ad}$, $\pi_{nd}$, $\pi_{dh}$ and $\pi_{dp}= 1 - \pi_{ad} - \pi_{nd} - \pi_{dh}$ be the respective subpopulation proportions; see  also Table \ref{Tab:PopStratDef}. 
 

\begin{table}[ht!]
	\centering
	\caption{\footnotesize Population stratification for semi-competing risks data. \label{Tab:PopStratDef}}
{\footnotesize	\begin{tabular}{c|c|c}
Name &  Stratum definition & Stratum proportion  \\
\hline
Always-diseased (ad) & $T_1(0)\le T_2(0)$ \& $T_1(1) \le T_2(1)$ & $\pi_{ad}$ \\[0.5em]
Never-diseased (nd) & $T_1(0) > T_2(0)$ \& $T_1(1) > T_2(1)$ & $\pi_{nd}$ \\[0.5em]
Disease harmed (dh) & $T_1(0) > T_2(0)$ \& $T_1(1) \le T_2(1)$ & $\pi_{dh}$ \\[0.5em]
Disease protected (dp) & $T_1(0) \le T_2(0)$ \& $T_1(1) > T_2(1)$ & $\pi_{dp}$ \\[0.5em]
	\end{tabular}}
\end{table}


Because $T_1(a)=\infty$ for some $a$ for the $dh$ and $dp$ strata,  well-defined causal effects that are of interest only exist within the $ad$ and the $nd$ strata, similar to cases where principal stratification has been utilized \citep{zhang2003estimation,shepherd2007sensitivity,ding2011identifiability,ding2017principal}. In the motivating example (Section \ref{Sec:Data}), the $ad$ and $nd$ proportions were estimated to be 35\% and 53\%, respectively, meaning nearly all the population belongs to a stratum within which there is a well-defined causal effect.    At each time $t$, we can define the following causal estimands
\begin{align}
	\label{Eq:estimand:ProbT2ad}
\Pr&[T_2(1)\le t\ |\ ad]-\Pr[T_2(0)\le t\ |\ ad],\\
	\label{Eq:estimand:ProbT2nd}
\Pr&[T_2(1)\le t\ |\ nd]-\Pr[T_2(0)\le t\ |\ nd],\\
	\label{Eq:estimand:ProbT1ad}
\Pr&[T_1(1)\le t\ |\ ad]-\Pr[T_1(0)\le t\ |\ ad].
\end{align}
Causal effects that better summarize the impact of $A$ could be based on the restricted mean survival time (RMST) \citep{chen2001causal}. For a given time point $t^\star$,  let $\mathcal{T}_j(a)=\min[T_j(a),t^\star], a,j=0,1$, 
 and define the  following additional causal estimands
\begin{align}
	\label{Eq:estimand:MeanT2ad}
E&[\mathcal{T}_2(1) - \mathcal{T}_2(0)\ |\ ad],\\
	\label{Eq:estimand:MeanT1ad}
E&[\mathcal{T}_1(1) - \mathcal{T}_1(0)\ |\ ad],\\
	\label{Eq:estimand:MeanT2afterT1ad}
E&\{[\mathcal{T}_2(1)-\mathcal{T}_1(1)] - [\mathcal{T}_2(0)-\mathcal{T}_1(0)]\ |\ ad\},\\
\label{Eq:estimand:MeanT2nd}
E&[\mathcal{T}_2(1) - \mathcal{T}_2(0)\ |\ nd].
\end{align}
Each of these estimands characterizes a different causal effect of $A$ on the outcomes within the $ad$ or $nd$ stratum. Effects \eqref{Eq:estimand:MeanT2ad}--\eqref{Eq:estimand:MeanT2afterT1ad} are defined within $ad$. In the motivating example, effect \eqref{Eq:estimand:MeanT2ad} is the mean (possibly negative) gain in life expectancy caused by APOE,  \eqref{Eq:estimand:MeanT1ad} is the effect of APOE on AD diagnosis age, and \eqref{Eq:estimand:MeanT2afterT1ad}, which is obtained by subtracting \eqref{Eq:estimand:MeanT1ad} from \eqref{Eq:estimand:MeanT2ad},   is the effect  on the residual lifetime with AD. If, for example, the latter equals zero, it means the same life expectancy with AD is expected for those with and without APOE within the $ad$ stratum. 
 Effect \eqref{Eq:estimand:MeanT2nd} is also of great interest, as it is the effect of APOE on life expectancy among the $nd$ population, thus it captures another form of APOE effect on survival.  Furthermore, contrasting 	\eqref{Eq:estimand:MeanT2ad} with \eqref{Eq:estimand:MeanT2nd} may provide information on the  heterogeneity of   APOE effect on survival. Of course, other  distribution summary measures (e.g. median time-to-events) can be used to define causal effects. 



\section{Identifiability}
\label{Sec:AssumpPrtIdent}
The focus of this section is on assumptions and identifiability.  Methods for estimation and inference under right censoring  are developed in Section \ref{Sec:EstInfer}.  Let $(A, T_1, T_2)$ be the actual treatment and event times. Throughout this paper,  we assume the following  two standard assumptions  \citep{hernan2019causal}.
\begin{assumption}
	\label{Ass:Sutva}
	\textit{Stable Unit Treatment Value Assumption (SUTVA)}.  $T_1=T_1(1)A + T_1(0)(1-A)$ and $T_2=T_2(1)A + T_2(0)(1-A)$.
\end{assumption}
\begin{assumption}
	\label{Ass:ignorabilityNoCens}
	\textit{Strong Ignorability}, $A \indep \{T_1(0),T_2(0)\},  \{T_1(1), T_2(1)\}$.
\end{assumption} 
In many observational studies that lack randomization, a conditional version of 	Assumption \ref{Ass:ignorabilityNoCens}, conditionally on covariates, is more plausible.

\subsection{Bounds}
\label{SubSec:bounds}
It is well-known that  principal stratum effects are not identified from the data, unless additional assumptions  are leveraged \citep{zhang2009likelihood,ding2011identifiability,tchetgen2014identification,ding2017principal}, and often bounds  are developed \citep{zhang2003estimation,long2013sharpening} .   In this section, we present various bounds for the causal effects based on the identifiable distribution of  $(A,T_1, T_2)$, constructed differently depending on the assumptions a researcher is willing to make. First, we present a new assumption,  tailored for semi-competing risks data. 
\begin{assumption}
\label{Ass:OrdPersev}
	\textit{Order preservation}. If $T_1(0)\le T_2(0)$, then $T_1(1)\le T_2(1)$.
\end{assumption}
In the motivating  example,  under Assumption \ref{Ass:OrdPersev}  anyone who would have been diagnosed with AD  without having APOE, would also have AD had they had APOE, but not necessarily at the same  age. It is  consistent with the overwhelming evidence  that APOE  significantly increases the risk of AD \citep{alzheimer20192019}. Assumption  \ref{Ass:OrdPersev} can be easily  modified for studies with a beneficing  effect of $A$ on the non-terminal event time.

 Assumption \ref{Ass:OrdPersev}  is less restrictive than the  often-invoked  monotonicity assumption \citep{zhang2003estimation,long2013sharpening}. 
Unlike  monotonicity, Assumption \ref{Ass:OrdPersev}  acknowledges the  realistic scenario that  $A$ may impact  survival time for either direction. Specifically, it allows that $T_2(0)-T_2(1)$ and $T_1(0) - T_1(1)$ are of different signs at the unit level. To see this, observe that $T_1(1)\le T_2(1)$ implies $T_2(0)-T_2(1) \le [T_2(0) - T_1(0)]  + [T_1(0) - T_1(1)]$. Finally,  Assumption \ref{Ass:OrdPersev} implies that the $dp$ stratum is empty, i.e., $\pi_{dp}=0$. 

As is typically the case in truncation-by-death problems, even under  Assumptions \ref{Ass:Sutva}--\ref{Ass:OrdPersev}, none of the causal effects \eqref{Eq:estimand:ProbT2ad}--\eqref{Eq:estimand:MeanT2nd} are non-parametrically identifiable from the data. These effects are however partially identifiable, in the sense that the data, coupled with the assumptions,  provide bounds on these effects.  For any event $\mathcal{Q}$, let $F_{j|\mathcal{Q}}(t)=\Pr(T_j\le t|\mathcal{Q})$, $S_{j|\mathcal{Q}}(t)=1-F_{j|\mathcal{Q}}(t)$ and $\eta_{\mathcal{Q}}=\Pr(T_1\le T_2|\mathcal{Q})$.  Under Assumptions \ref{Ass:Sutva}--\ref{Ass:OrdPersev}, $\Pr(T_2(0)\le t\ |\ ad), \Pr(T_2(1)\le t\ |\ nd)$ and $\Pr(T_1(0)\le t\ |\ ad)$ are identified from the data, while $\Pr(T_2(1)\le t\ |\ ad), \Pr(T_2(0)\le t\ |\ nd)$ and $\Pr(T_1(1)\le t\ |\ ad)$  are only partially identified. The following proposition establishes partial identifiability under Assumptions \ref{Ass:Sutva}--\ref{Ass:OrdPersev}.
\begin{proposition}
	\label{Prop:PartIdent}
	Under Assumptions \ref{Ass:Sutva}--\ref{Ass:OrdPersev},  effects \eqref{Eq:estimand:ProbT2ad}--\eqref{Eq:estimand:ProbT1ad} are partially identified by
		\begin{align}
\label{Eq:PartIdentT2ad}	
	& \mathcal{L}_{2,ad}(t) \le \Pr[T_2(1) \le t |ad]  -	\Pr[T_2(0)\le t |ad] \le  \mathcal{U}_{2,ad}(t), \\	
	 \label{Eq:PartIdentT2nd}
	& \mathcal{L}_{2,nd}(t) \le \Pr[T_2(1)\le t|nd]  - \Pr[T_2(0)\le t|nd] \le  \mathcal{U}_{2,nd}(t), \\	
	\label{Eq:PartIdentT1ad}
	& \mathcal{L}_{1,ad}(t) \le \Pr[T_1(1)\le t|ad]  - \Pr[T_1(0)\le t|ad] \le  \mathcal{U}_{1,ad}(t), 
	\end{align}
	\end{proposition}
	\noindent \textit{where}
	\begin{align}
	\begin{split}
		\label{Eq:Bounds}
	& \mathcal{L}_{2,ad}(t) =  \max\left\{0\ , 1-\frac{S_{2|A=1}(t)}{\eta_{A=0}}\right\} - F_{2|A=0, T_1 \le T_2}(t), \\
	& \mathcal{U}_{2,ad}(t) = \min\left\{1\ , F_{2|A=1}(t)\frac{\eta_{A=1, T_2\le t}}{\eta_{A=0}}\right\} - F_{2|A=0, T_1 \le T_2}(t), \\
		& \mathcal{L}_{2,nd}(t) =  F_{2|A=1, T_1 > T_2}(t) - \min\left\{1\ , F_{2|A=0}(t)\frac{1-\eta_{A=0,T_2\le t}}{1-\eta_{A=1}}\right\},\\
		& \mathcal{U}_{2,nd}(t) =  F_{2|A=1, T_1 > T_2}(t) - \max\left\{0\ , 1-\frac{S_{2|A=0}(t)}{1-\eta_{A=1}}\right\},\\
		& \mathcal{L}_{1,ad}(t) =  \max\left\{0\ , 1-\frac{S_{1|A=1}(t)}{\eta_{A=0}}\right\} - F_{1|A=0, T_1 \le T_2}(t), \\
		& \mathcal{U}_{1,ad}(t) = \min\left\{1\ , \frac{F_{1|A=1}(t)}{\eta_{A=0}}\right\} - F_{1|A=0, T_1 \le T_2}(t).
		\end{split}
		\end{align}
	The proof is given in Section A.1 of  the Supplementary Materials (SM).  From the proof it could also be seen that under Assumptions \ref{Ass:Sutva}--\ref{Ass:OrdPersev}, the strata proportions are identified by 	$\pi_{ad}=\eta_{A=0}$,
	$\pi_{dh}=\eta_{A=1}-\eta_{A=0}$, $\pi_{nd} = 1- \eta_{A=1}$. 
	 Since $T_1(a)$ and $T_2(a)$ are positive with probability one, RMST effects \eqref{Eq:estimand:MeanT1ad}--\eqref{Eq:estimand:MeanT2nd} are  (partially) identified by recalling that for any non-negative random variable, $E(T)=\int_{0}^{\infty}[1-\Pr(T\le t)]dt$.   
	 
Proposition \ref{Prop:PartIdent} provides bounds for the different causal effects. Had we had unlimited data, calculating the bounds would have been simple. In the absence of right censoring, we could have simply calculate $F_{2|A=a}(t), F_{2|A=a, T_1 \le T_2}(t)$, $\eta_{A=a}, \eta_{A=a,T_2\le t}$ from the data. As for $F_{1|A=a}(t)$, we could have used $F_{1|A=a}(t)=\eta_{A=a}F_{1|A=a,T_1\le T_2}(t)$.

Generally, bounds derived under additional assumptions may  provide narrower bounds \citep{zhang2003estimation,long2013sharpening}. Of course, these  assumptions need to be plausible in  the scientific problem at hand for the bounds to be valid. One particular bound that may not be informative enough is $\mathcal{L}_{1,ad}(t)$. As $t$ increases, the second term $F_{1|A=0,T_1\le T_2}(t)$ get closer to one while the first term is not, because $\Pr(T_1 < \infty|{A=1})<1$. As a remedy for this problem, consider the following assumption and proposition providing a more informative lower bound for \eqref{Eq:estimand:ProbT1ad}.
\begin{assumption}
	\label{Ass:Rank} For a given $t$,  $\Pr[T_1(1) \le t| ad] \ge \Pr[T_1(1) \le t| dh]$.
\end{assumption}
This assumption resembles the ranked average score assumption \citep{zhang2003estimation}, but tailored for time-to-event data. It means that the probability of an always-diseased person to be diagnosed with the disease by  time $t$ (under $A=1$) is larger than the probability of same event for a disease-harmed person (also under $A=1$). In practice, Assumption \ref{Ass:Rank}  can hold for some, all, or none of the values of $t$.  It will hold if those in the $ad$ stratum tend to be diagnosed at earlier ages than those in the $dh$ stratum.  
\begin{proposition}
	\label{Prop:PartIdentRank}
	Under Assumptions \ref{Ass:Sutva}--\ref{Ass:OrdPersev},  for any $t$ for which Assumption \ref{Ass:Rank} holds, $\Pr[T_1(1)\le t|ad]  - \Pr[T_1(0)\le t|ad]$ is partially identified by
	$$
	 \widetilde{\mathcal{L}}_{1,ad}(t) \le \Pr[T_1(1)\le t|ad]  - \Pr[T_1(0)\le t|ad] \le  \mathcal{U}_{1,ad}(t)
	 $$
	 where $\widetilde{\mathcal{L}}_{1,ad}(t)=F_{1|A=1 T_1 \le T_2}(t) - F_{1|A=0, T_1 \le T_2}(t)$ and $\mathcal{U}_{1,ad}(t)$ is given in Proposition \ref{Prop:PartIdent}.
\end{proposition}
The proof is given in Section A.2 of the SM.  In some studies, the obtained bounds, with or without additional assumptions, might be too wide to have practical utility. One approach is to improve the bounds by leveraging an additional covariate. Following \cite{long2013sharpening},  we develop below narrower bounds  utilizing a pre-intervention discrete covariate $Z$. In the motivating example, $Z$ can be, for example, gender.  An adjusted version of Proposition \ref{Prop:PartIdent}  can be obtained under the following   assumption.
\begin{assumption}
	\label{Ass:ignorabilityZ}
	$A \indep Z, \{T_1(0),T_2(0)\}, \{T_1(1), T_2(1)\}$.
\end{assumption}
For example, we  show in Section A.3 of the SM that under Assumptions \ref{Ass:Sutva},  \ref{Ass:OrdPersev} and \ref{Ass:ignorabilityZ}, adjusted bounds for \eqref{Eq:estimand:ProbT2ad} are given by $\mathcal{L}^{Z}_{2,ad}(t),\mathcal{U}^{Z}_{2,ad}(t)$, where
\begin{align}
\begin{split}
\label{Eq:AdjustedBound}
& \mathcal{L}^{Z}_{2,ad}(t) =  \sum_{z}\nu(z)\left[\max\left\{0\ , 1-\frac{S_{2|A=1,Z=z}(t)}{\eta_{A=0,Z=z}}\right\} - F_{2|A=0, T_1 \le T_2,Z=z}(t)\right], \\
& \mathcal{U}^{Z}_{2,ad}(t) =  \sum_{z}\nu(z)\left[\min\left\{1\ , F_{2|A=1,Z=z}(t)\frac{\eta_{A=1,T_2\le t, Z=z}}{\eta_{A=0,Z=z}}\right\} - F_{2|A=0, T_1 \le T_2,Z=z}(t)\right],
\end{split}
\end{align}
and  $\nu(z)=\Pr(Z=z|A=0, T_1\le T_2)$. Furthermore, the following proposition establishes that the adjusted bounds are always within the unadjusted bounds.
\begin{proposition}
	\label{Prop:AdjustedWithin}
	Under Assumptions \ref{Ass:Sutva},  \ref{Ass:OrdPersev} and \ref{Ass:ignorabilityZ}, $\mathcal{L}_{2,ad}(t)\le \mathcal{L}^{Z}_{2,ad}(t)$ and $\mathcal{U}_{2,ad}(t)\ge  \mathcal{U}^{Z}_{2,ad}(t)$.
\end{proposition}
The proof is given in Section A.4 of the SM. Proposition \ref{Prop:AdjustedWithin} is the analogue of Proposition 1 in \cite{long2013sharpening}, who developed bounds when the outcome is binary, and also investigated under which conditions the statement in the inequalities can be made in the sharp sense. Of note is that due to finite-sample variation,  Proposition \ref{Prop:AdjustedWithin} may not hold empirically. Therefore, as suggested by \cite{long2013sharpening},  the bounds $\max\{\mathcal{L}_{2,ad}(t), \mathcal{L}^{Z}_{2,ad}(t)\}$ and $\min\{\mathcal{U}_{2,ad}(t),  \mathcal{U}^{Z}_{2,ad}(t)\}$ can be used.

\begin{figure}[h!]
	\centering
	\includegraphics[scale = 0.35]{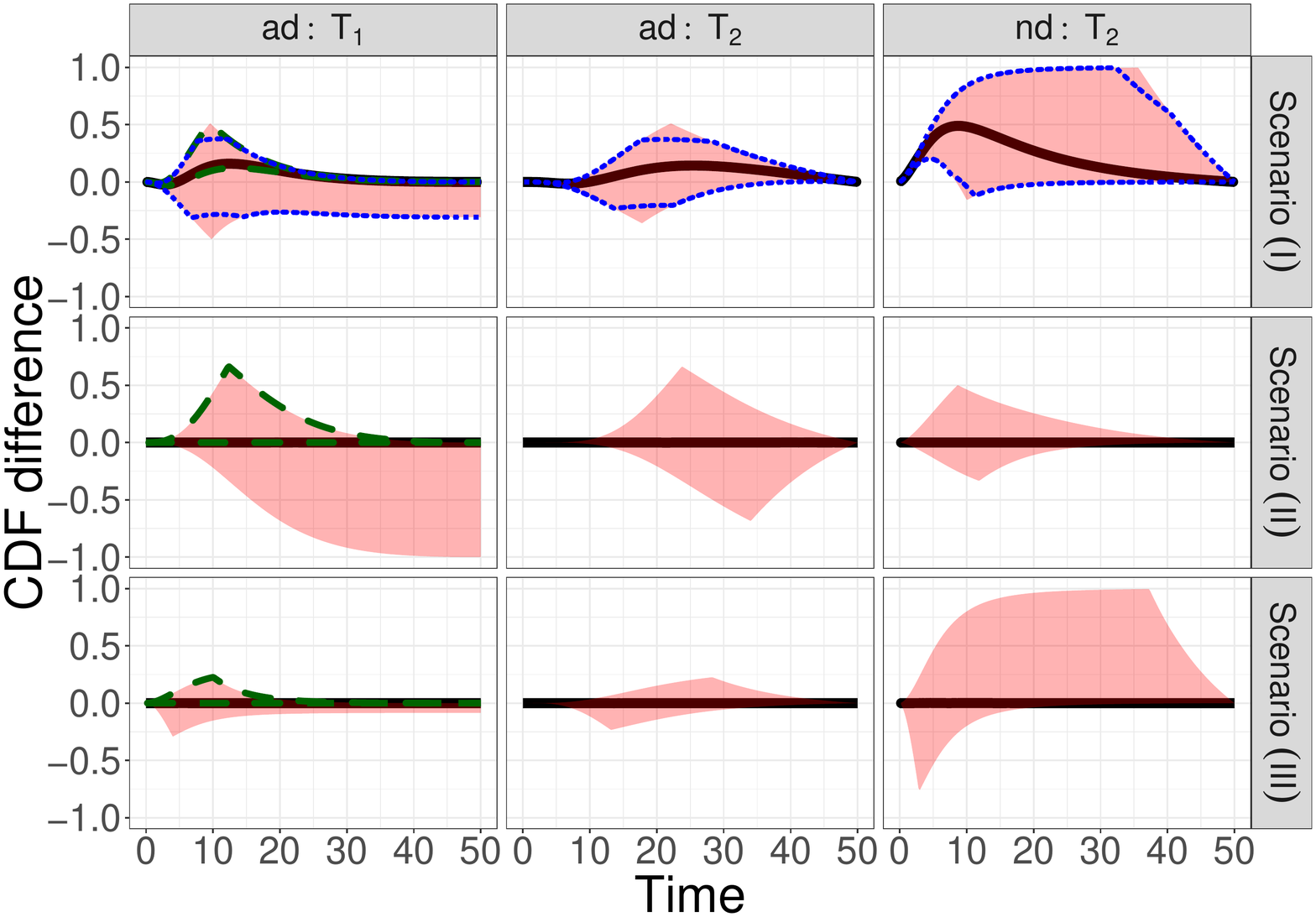}
	\caption{\footnotesize True causal effects  \eqref{Eq:estimand:ProbT2ad}--\eqref{Eq:estimand:ProbT1ad} and different bounds under three simulated scenarios.}
	\label{Fig:LargeSample}
\end{figure}

We summarize this section by an illustrative comparison of the various bounds. Figure \ref{Fig:LargeSample} presents the bounds for three scenarios; the data-generating  mechanism (DGM) and the resulting per-stratum cumulative distribution functions are described in Section D.1 of the SM.  Under Scenario (I) (top row), Assumptions \ref{Ass:OrdPersev} and \ref{Ass:Rank} hold, and $A=1$ shorten the $ad$ time-to-disease and time-to-death,  and $nd$ die faster under $A=1$.  
The bounds $\{\mathcal{L}_{1,ad}(t), \mathcal{U}_{1,ad}(t)\}$, $\{\mathcal{L}_{2,ad}(t),  \mathcal{U}_{2,ad}(t)\}$,  and $\{\mathcal{L}_{2,nd}(t), \mathcal{U}_{2,nd}(t)\}$  (in pink shade) are quite wide in this scenario. The bound 	 $\widetilde{\mathcal{L}}_{1,ad}(t)$ (dashed green line), derived in Proposition \ref{Prop:PartIdentRank} is far more informative, being only slightly lower than the true difference (black solid curve). The   adjusted bounds (blue dotted lines)  were narrower than the unadjusted bounds, most notably for $t$ values for which the unadjusted bounds are very wide. For the causal effect  on time-to-disease within the $ad$ (top left corner)  the narrower bound is obtained as $\{\widetilde{\mathcal{L}}_{1,ad}(t), \mathcal{U}^{Z}_{2,ad}(t)\}$.  The bounds for time-to-death among the $nd$ are quite wide. This is likely because there is least information on this group  from the observed data, as the true strata probability were $(\pi_{ad}, \pi_{nd},  \pi_{dh})= (0.49, 0.15, 0.36)$.

 To further investigate this point,  consider the second and third rows of  Figure \ref{Fig:LargeSample}, both under null effects with no covariate $Z$ available. In Scenario (II), the $nd$ stratum comprised most of the population $(\pi_{ad}, \pi_{nd},  \pi_{dh})= (0.13, 0.59, 0.28)$ and the bounds were quite narrow for  the within $nd$ effect $\eqref{Eq:estimand:ProbT2nd}$, and less informative for the $ad$. In Scenario (III), when $\pi_{ad}$ was substantial $(\pi_{ad}, \pi_{nd},  \pi_{dh})= (0.70, 0.07, 0.23)$, the bounds were quite narrow for the within $ad$ causal effects $\eqref{Eq:estimand:ProbT2ad}$ and $\eqref{Eq:estimand:ProbT1ad}$, but not so informative for the $nd$ stratum. In conclusion, if the proportion of one of the stratum $ad$ or $nd$ is large enough, the bounds can be useful, and a covariate $Z$ may help to make the bounds narrower. If Assumption \ref{Ass:Rank} is plausible, the lower bound $\widetilde{\mathcal{L}}_{1,ad}(t)$ can be quite informative.


 A covariate that can be used to sharpen the bounds is not always available, or even if it is,  the  bounds might still be too wide. Therefore,  alternative strategies for an analysis are desirable. In Section B of the SM we describe a sensitivity analysis approach arising from the proof of Proposition \ref{Prop:PartIdent}. However, our main approach towards conducting sensitivity analysis for casual effect estimation from semi-competing risks data stems from the ubiquitous  illness-death modeling of semi-competing risks data.
 

\subsection{Identification by frailty assumptions}
\label{SubSec:IdentFrail}
We turn to present stronger assumptions under which effects \eqref{Eq:estimand:ProbT2ad}--\eqref{Eq:estimand:ProbT1ad}  are identified from the data. The proposed overarching  strategy adapts the illness-death model so causal estimands can be captured by the model, by formulating two illness-death models and tying them together via  a bivariate frailty random variable $\bgamma=(\gamma_0,\gamma_1)$.   For  $a=0,1$, let 
\begin{align*}
\lambda_{01}(t|a,\bgamma)&=\lim\limits_{\Delta\rightarrow0}\Delta^{-1}\Pr[T_1\in [t,t+\Delta)|A=a,T_1 \ge t,T_2\ge t,\bgamma], \quad t>0,\\ \lambda_{02}(t|a,\bgamma)&=\lim\limits_{\Delta\rightarrow0}\Delta^{-1}\Pr[T_2\in [t,t+\Delta)|A=a,T_1 \ge t,T_2\ge t,\bgamma],\quad t>0, \\ \lambda_{12}(t|a,t_1,\bgamma)&=\lim\limits_{\Delta\rightarrow0}\Delta^{-1}\Pr[T_2\in [t,t+\Delta)|A=a,T_1=t_1,T_2\ge t,\bgamma], \quad t>t_1>0
\end{align*}
be the three cause-specific hazard functions under the semi-competing risks setting \citep{xu2010statistical}, associated with the transitions $j\rightarrow k, jk=01,02,12$  (0: healthy, 1: disease, 2: death);  see Figure C.1 in  the SM for illustration.  Our approach requires assumptions on the conditional (in)dependence of $\{T_1(0),T_2(0)\}$ and $\{T_1(1),T_2(1)\}$ given $\bgamma$. 
\begin{assumption}
	\label{Ass:FrailtyForIdent} 
	There exists a bivariate random variable $\bgamma=(\gamma_0,\gamma_1)$ such that 
	\begin{enumerate}[(i)]
		\item  Given $\bgamma$, the joint distribution of the potential event times can be factored as follows 
		\begin{equation}
		f[T_1(0),T_2(0),T_1(1), T_2(1)|\bgamma] = f[T_1(0),T_2(0)|\gamma_0]f[T_1(1),T_2(1)|\gamma_1],
		\end{equation}
		where $f(\cdot)$ denotes a density function of a possibly-multivariate random variable. 
\item 	 $A \indep \{T_1(0),T_2(0),\gamma_0\},  \{T_1(1), T_2(1),\gamma_1\}$.
\item The frailty variable operates multiplicatively on the hazard functions. That is,  $\lambda_{jk}(t|a,\bgamma)=\gamma_a\widetilde{\lambda}_{jk}(t|a)$ for $jk=01, 02, a= 0,1$ and $\lambda_{12}(t|a,t_1,\bgamma)=\gamma_a\widetilde{\lambda}_{12}(t|a,t_1)$ for $a=0,1$, for some $\widetilde{\lambda}_{jk}$ functions.
			\item The probability density function of $\bgamma$, $f_{\btheta}(\bgamma)$, is known up to a finite dimensional  parameter $\btheta$ that is identifiable from the observed data distribution.
	\end{enumerate}
\end{assumption}
Part $(i)$  of Assumption \ref{Ass:FrailtyForIdent} implies a cross-world independence conditionally on the unobserved $\bgamma$. This assumption is more general than Assumption 4 in \cite{comment2019survivor} in that their assumption additionally assumes $\gamma_0=\gamma_1$. However, $\gamma_0=\gamma_1$ would generally not hold unless all treatment effect modifiers can be measured and correctly accounted for via changes in the $\widetilde{\lambda}_{jk}$ functions between $a=0$ and $a=1$.     Part $(ii)$ is just an adaption of Assumption \ref{Ass:ignorabilityNoCens} for the frailty case  and  $(iii)$ and $(iv)$ are standard assumptions  for identification of the observed data distribution.   The strategy of how to model the distribution of $\bgamma$ dictates two forms of dependence, cross-world and within-world.  We give below three examples of  frailty distributions specifications. 
\begin{example}
	A bivariate normal distribution with mean  zero and a given  correlation $\rho$ is assumed for $[\log(\gamma_1),\log(\gamma_2)]$. The variances $Var(\log{\gamma_a})$ are identifiable from the data.
\end{example}
\begin{example}
	The frailty variables $(\gamma_0, \gamma_1)$ are independent Gamma random variables, with mean one and variances $\theta_0$ and $\theta_1$, respectively, both identifiable from the data.
\end{example}
\begin{example}
	The frailty variables are $(\gamma_0, \gamma_1)$ correlated Gamma variables, with mean one,  variances $\theta_0$ and $\theta_1$, and correlation $\rho$. The frailty variances $\theta_0$ and $\theta_1$ are identifiable from the data, while  $\rho$ is unidentifiable  and need to be supplied by the researcher.
\end{example}
 Crucially, Example 3 entails  a weaker assumption than  assuming that $\gamma_0=\gamma_1$. The latter implies the same dependence between the non-terminal and terminal event times in both worlds. Thus,     it constraints the effect of $A$ on $\{T_1,T_2\}$,  and hence it constitutes a strong assumption, that should not hold at least in some settings.  Assumption \ref{Ass:FrailtyForIdent}, especially its first part, is generally a strong assumption. Nevertheless, it opens the way for causal interpretation of existing semi-competing risks models. 
\begin{proposition}
	\label{Prop:IdentFrail}
Under Assumptions \ref{Ass:Sutva} and \ref{Ass:FrailtyForIdent}, the causal effects \eqref{Eq:estimand:ProbT2ad}--\eqref{Eq:estimand:ProbT1ad}	are identified by
		\begin{align*}
\Pr&[T_2(1) \le t |ad]  -	\Pr[T_2(0)\le t|ad] \\
&= \int_{0}^{\infty}\int_{0}^{\infty}[F_{2|T_1 \le T_2, A=1,\gamma_1}(t) - F_{2|T_1 \le T_2, A=0,\gamma_0}(t)\}] \frac{\eta_{A=0,\gamma_0}\eta_{A=1,\gamma_1}f_{\btheta}(\bgamma)d\bgamma}{\int_{0}^{\infty}\int_{0}^{\infty}\eta_{A=0,\gamma'_0}\eta_{A=1,\gamma'_1}f_{\btheta}(\bgamma')d\bgamma'},\\	 
\Pr&[T_2(1) \le t|nd]  - \Pr[T_2(0)\le t|nd] \\ 
&=\int_{0}^{\infty} \int_{0}^{\infty}[F_{2|T_1 > T_2, A=1,\gamma_1}(t) - F_{2|T_1 > T_2, A=0,\gamma_0}(t)] \frac{(1 -\eta_{A=0,\gamma_0})(1 - \eta_{A=1,\gamma_1})f_{\btheta}(\bgamma)d\bgamma}{\int_{0}^{\infty}\int_{0}^{\infty}(1 - \eta_{A=0,\gamma'_0})(1 - \eta_{A=1,\gamma'_1})f_{\btheta}(\bgamma')d\bgamma'},\\	
 \Pr&[T_1(1)\le t|ad]  - \Pr[T_1(0)\le t|ad] \\ 
&=\int_{0}^{\infty} \int_{0}^{\infty}[F_{1|T_1 \le T_2, A=1,\gamma_1}(t) - F_{1|T_1 \le T_2, A=0,\gamma_0}(t)\}] \frac{\eta_{A=0,\gamma_0}\eta_{A=1,\gamma_1}f_{\btheta}(\bgamma)d\bgamma}{\int_{0}^{\infty}\int_{0}^{\infty}\eta_{A=0,\gamma'_0}\eta_{A=1,\gamma'_1}f_{\btheta}(\bgamma')d\bgamma'},
\end{align*}
where for any event $\mathcal{Q}$, $F_{j|\mathcal{Q},\bgamma}(t)= \Pr(T_j \le t|\mathcal{Q},\bgamma)$. Furthermore, all the integrals  can be consistently estimated from the data.
\end{proposition}
The proof is given in Section A.5 of the SM. RMST-like estimands \eqref{Eq:estimand:MeanT1ad}--\eqref{Eq:estimand:MeanT2nd} are  identified by suitable integration over $t$. A sensitivity analyses based on Proposition \ref{Prop:IdentFrail} can be carried out by repeating the analysis (e.g. under Example 3) for different $\rho$ values. 

Regarding the choice of which frailty distribution to use in practice, it is well-known that misspecification of the frailty distribution leads to only small bias in the estimated cumulative incidence function, integrated over the frailty distribution; see \cite{gorfine2020marginalized} and references therein.

\begin{figure}[h]
	\centering
	\includegraphics[width = 0.5\linewidth]{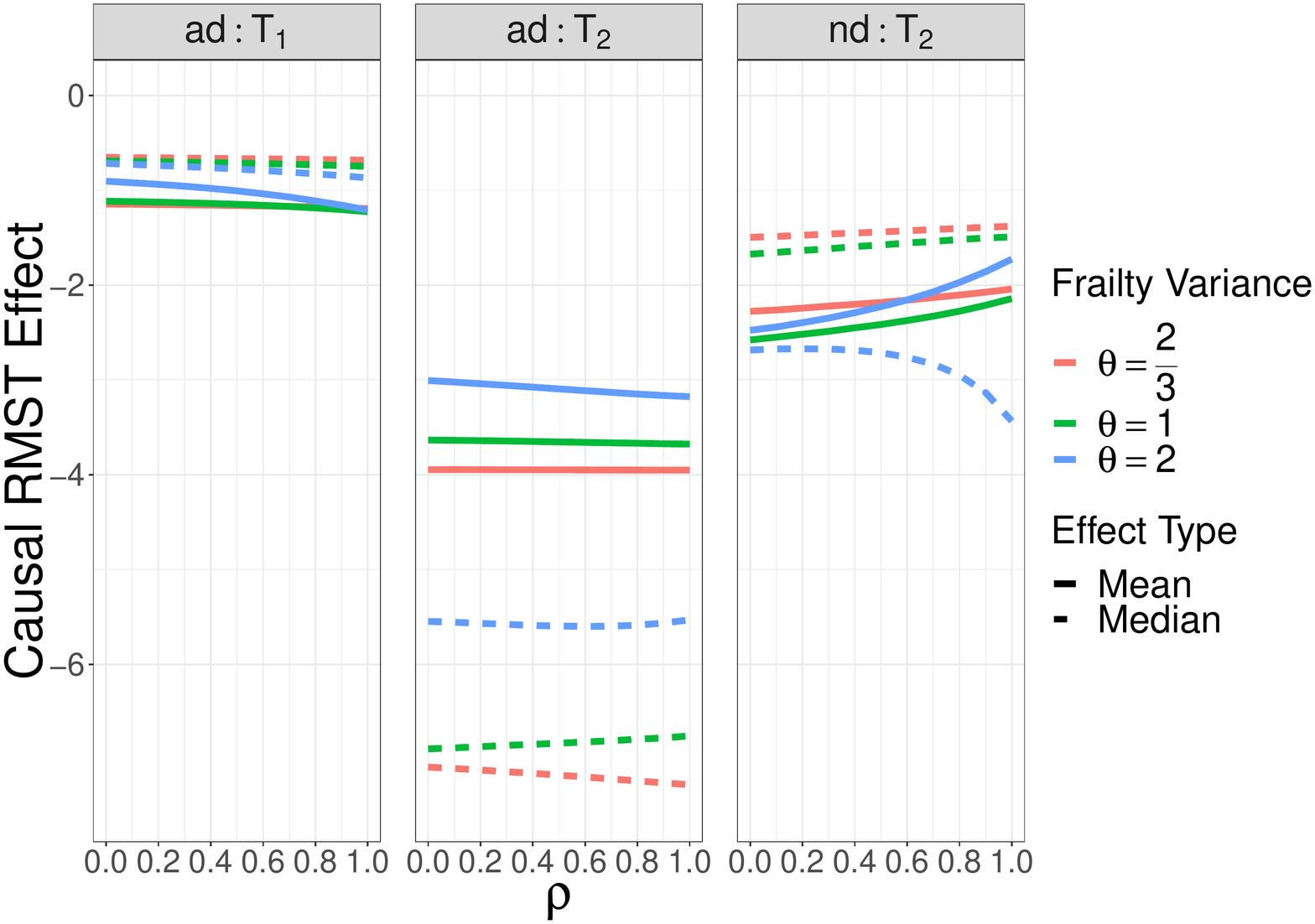}
\caption{\footnotesize Causal mean and median RMST effects as a function of the frailty variance $\theta=\theta_0=\theta_1=2/3, 1, 2$, (corresponding to Kendall's $\tau=1/4,1/3,1/2$) and correlation between the frailty variables $\rho \in [0,1]$. RMST effects were calculated at $t^\star=15$.}
\label{Fig:FrailtyCausalEffects}
\end{figure}

Figure \ref{Fig:FrailtyCausalEffects} illustrates the  causal RMST effects \eqref{Eq:estimand:MeanT2ad}, \eqref{Eq:estimand:MeanT1ad}, and \eqref{Eq:estimand:MeanT2nd}  under  the DGM  described in Section  \ref{Sec:EstInfer} and Section D.1 of the SM, and the frailty specification given in Example 3, with equal variances $\theta=\theta_0=\theta_1$. Stratum proportions varied between 41\%--45\% for $ad$ stratum and 8\%--12\% for the $nd$ stratum. The causal RMST effects, calculated at $t^\star=15$,   were  sensitive to $\rho$ when the stratum proportion was very small and the frailty variance was large, and were otherwise quite robust to $\rho$.  

\subsection{Right censoring}
\label{Sec:Cens}

Time-to-event outcome data are often  not fully observed for all individuals due to loss to follow-up (i.e., right censoring). Two issues arise when considering  right censoring.  The first is identification of the causal estimand, considered in this section, and the second is how to estimate the different quantities  and provide inference, which is embedded in the non-parametric and semi-parametric methods  presented in Section \ref{Sec:EstInfer}. Let $C(a)$ be the censoring time under $A=a$.  We first adjust Assumption \ref{Ass:ignorabilityNoCens} for the case of censoring.
\begin{assumption}
	\label{Ass:ignorabilityCens}
	\textit{Strong Ignorability} $A \indep \{T_1(0),T_2(0),C(0)\}, \{T_1(1),T_2(1),C(1)\}$.
\end{assumption}
Our next assumption is a within-world independent-censoring assumption.
\begin{assumption}
	\label{Ass:IndepCens}
	\textit{Independent censoring} $\{T_1(a), T_2(a)\} \indep C(a)$ for $a=0,1$. 
\end{assumption}
Assumption	\ref{Ass:IndepCens} asserts that, under each treatment value,  the two event times and censoring time are independent. We also assume the standard assumption that there exists an end of follow-up time $\overline{c}$ and that $Pr(T_j(a)>\overline{c}) > 0$ for $j=1,2$.   Combining Assumptions 	\ref{Ass:ignorabilityCens} and  	\ref{Ass:IndepCens} together with Assumption \ref{Ass:Sutva}, we get the standard independent censoring assumption for semi-competing risks data  $T_1,T_2\indep C$, with $C$ being the censoring time under $A$. 


\subsection{Covariates inclusion}
\label{Sec:Covar}

 There are number of reasons to include baseline covariates $\bX$ in our setting. First, because in observational studies  the intervention is not randomized, Assumption \ref{Ass:ignorabilityCens} is more plausible when made conditionally on $\bX$. Second,  the independent censoring assumption (Assumption \ref{Ass:IndepCens}) is also more plausible when made conditionally on $\bX$ \citep{andersen1993statistical}. Assumption \ref{Ass:FrailtyForIdent} is also more plausible when made conditionally on covariates. For simplicity of presentation, we use the same vector $\bX$  for  all assumptions to hold. In Section A.6 of the SM, we adapt  Propositions \ref{Prop:PartIdent} and \ref{Prop:IdentFrail} to include covariates.  An additional reason to include covariates is to improve estimators' efficiency.  In the motivating example (Section \ref{Sec:Data}), we use gender and  race as  covariates. Note that the role of $\bX$ is different from the role of the covariate $Z$ we utilized in Section \ref{SubSec:bounds}.
 
 While a variety of statistical models have been used for semi-competing risks data, how to elucidate information on causal effects is typically overlooked and,  in practice, researchers often focus on specific coefficients in regression models. Therefore, a main motivation for us to consider models is to enable translation of existing  model results into knowledge on the causal effects of interest.   \cite{comment2019survivor} used parametric models coupled with Bayesian methods for estimating the TV-SACE and RM-SACE. We focus on frequentist estimation of and inference about  causal effects \eqref{Eq:estimand:MeanT2ad}--\eqref{Eq:estimand:MeanT2nd} under semi-parametric frailty models that  align with Assumption \ref{Ass:FrailtyForIdent}. Details  are given in Section \ref{SubSec:semiparam}.


\section{Estimation and Inference}
\label{Sec:EstInfer}
So far, our focus has been on definitions of causal estimands,  assumptions and their plausibility, and identification strategies. We now turn to estimation of the identifiable components based on  time-to-event right-censored semi-competing risks data, and carrying out inference for these quantities.  We first propose non-parametric estimators and discuss their asymptotic properties, before considering semi-parametric models. For the non-parametric estimators, our efforts are focused on the quantities  $\eta_{A=a}$, $\eta_{A=a, T_2\le t}$,  $S_{1|A=a}$, $S_{2|A=a}$, $S_{1|A=a,T_1\le T_2}$, $S_{2|A=a,T_1\le T_2}$, and  $S_{2|A=a,T_1 > T_2}$ (for $a=0,1$). These are the building blocks of the bounds. For the semi-parametric estimation, we focus on a frailty proportional hazard model that allows estimation of  causal effects utilizing Proposition 	\ref{Prop:IdentFrail}.  

For each participant $i$, $i=1,...,n$ denote by $Z_i$ a covariate that may be used for obtaining sharper bounds (when available),  by $\bX_i$ a vector of  baseline covariates (when available), by   $A_i\in \{0,1\}$, the intervention value, and by $T_{ij}$ the  non-terminal ($j=1$) and terminal ($j=2$) event times. Let also $C_i$ be the censoring time. The observed data for each unit is $(Z_i, \bX_i, A_i, \widetilde{T}_{i1},\delta_{i1}, \widetilde{T}_{i2}, \delta_{i2})$  
where $\widetilde{T}_{i1}=\min(T_{i1},T_{i2},C_i)$ and $\widetilde{T}_{i2}=\min(T_{i2},C_i)$ are the observed times,  and $\delta_{i1}=I\{\widetilde{T}_{i1}=T_{i1}\}$  and $\delta_{i2}=I\{\widetilde{T}_{i2}=T_{i2}\}$ are the event  indicators of disease and death, respectively. Let also $Y_{ij}(t)=I\{t \le \widetilde{T}_{ij}\}$ be the at-risk  processes.

\subsection{Non-parametric Estimation}
 \label{SubSec:nonparam}

Starting with  $S_{2|A=a}(t)$, it can be  simply estimated  by  the Kaplan-Meier (KM) estimator of $T_2$ within the $\{A=a\}$ group. 
Regarding $S_{1|A=a}(t)$, observe that
\begin{equation}
\label{Eq:S1est}
S_{1|A=a}(t)=Pr(T_1>t|A=a)
= -\int_{0}^{\infty}S_{1|A=a,T_2=s}[\min(t,s)]dS_{2|A=a}(s),
\end{equation}
because for $t>s$, $S_{1|A=a,T_2=s}(t)=S_{1|A=a,T_2=s}(s)$, i.e., among those who die at time $s$, no new disease cases are possible later than $s$.   The function $S_{1|A=a,T_2=t}$ is well-defined,  and thus can be estimated non-parametrically. An explicit model for  $S_{1|A=a,T_2=t}$  is of little interest, as it is hard to motivate a model for the  nonstandard distribution of $T_1|T_2=t$.  

We propose to estimate  $S_{1|A=a,T_2=t}(t)$   by a kernel-smoothed  KM estimator \citep{beran1981nonparametric} within  the $\{A=a,\delta_2=1\}$ group. Estimation among this group is valid because under Assumption \ref{Ass:IndepCens}, $S_{1|A=a,T_2=t}(t)=S_{1|A=a,T_2=t,\delta_2=1}(t)$. Let $K(\cdot)$ be a  kernel function ($K(t)\ge0$, $\int K(t)dt=1$), and let $h_n\rightarrow0$  be a sequence of bandwidths. In practice, one might  consider a separate bandwidth for each treatment level. For simplicity of presentation, we assume $h_n$ is the same for both $a=0,1$. Let
 $Q(t,a;h_n)=\sum_{i=1}^{n}I\{A_i=a\}\delta_{i2}K[(t-T_{i2})/h_n]$, and $W_i(t,a;h_n)=I\{A_i=a\}\delta_{i2}K[(t-T_{i2})/h_n]/Q(t,a;h_n)$. The kernel KM estimator for $S_{1|A=a,T_2=t}(t)$  is then defined by
 \begin{equation*}
\widehat{S}_{1|A=a,T_2=t}(t)=\prod_{i=1}^{n}\left[1-\frac{W_i(t,a;h_n)}{\sum_{j=1}^{n}Y_{1j}(\widetilde{T}_{i1})W_j(t,a;h_n)}\right]^{\delta_{i1}\delta_{i2}I\{A_i=a, \widetilde{T}_{i1}\le t\}}.
 \end{equation*}
We propose to estimate $\eta_{A=a}$ in similar fashion utilizing that 
 \begin{align}
 \begin{split}
 \label{Eq:etaAest}
 \eta_{A=a}=Pr(T_1\le T_2|A=a)= -\int_{0}^{\infty}F_{1|A=a,T_2=t}(t)dS_{2|A=a}(t).
 \end{split}
 \end{align}
Turning to the other components, by  calculations given in Section C.1 of the SM,
\begin{align}
\label{Eq:etaAestT}
\eta_{A=a,T_2\le t} &=  -\frac{1}{F_{2|A=a}(t)}\int_{0}^{t}F_{1|A=a,T_2=s}(s)dS_{2|A=a}(s),\\
S_{1|A=a, T_1 \le T_2}(t)  &=-\frac{1}{\eta_{A=a}}\int_{t}^{\infty}[S_{1|A=a,T_2=s}(t)-S_{1|A=a,T_2=s}(s)]dS_{2|A=a}(s),\\ \label{Eq:S2AT1lT2est}
S_{2|A=a, T_1 \le T_2}(t) &= 1 - \frac{\eta_{A=a, T_2\le t}F_{2|A=a}}{\eta_{A=a}},\\ \label{Eq:S2AT1hT2est}
S_{2|A=a, T_1 > T_2}(t) &= 1 - \frac{(1 - \eta_{A=a, T_2\le t})F_{2|A=a}}{1 - \eta_{A=a}}.
\end{align}
Thus, estimators  $\widehat{S}_{1|A=a}(t)$, $\widehat{\eta}_{A=a}$, $\widehat{\eta}_{A=a,T_2\le t}$, $\widehat{S}_{2|A=a, T_1 \le T_2}(t)$ and $\widehat{S}_{2|A=a, T_1 > T_2}(t)$, are obtained by substituting the  KM and smoothed KM estimators into  \eqref{Eq:S1est}--\eqref{Eq:S2AT1hT2est}.  Covariate-adjusted bounds \eqref{Eq:AdjustedBound} are estimated  at each level of $Z$, and then averaged using $\widehat{\nu}(z)= \widehat{\Pr}(Z=z) \widehat{\eta}_{Z=z}/\widehat{\Pr}(T_1\le T_2)$. Note, because the estimators are risk-set based, they accommodate not only right censoring but also delayed entry, as we illustrate  in Section \ref{Sec:Data}. 
Our \textbf{R} package \texttt{CausalSemiComp} implements methodology presented in this Section utilizing the \texttt{prodlim} package \citep{prodlim} for calculating the smoothed KM estimators.

The asymptotic properties of the KM estimator  and of the smoothed KM estimator are well established. For example, under Assumption \ref{Ass:IndepCens} and additional regularity assumptions, the smoothed KM estimator is consistent and weakly converges to a Gaussian process \citep{beran1981nonparametric,dabrowska1987non}. A sketch of the proof of the consistency of  our proposed estimators is provided in Section C.2 of the SM.

\subsection{Semi-parametric frailty models}
 \label{SubSec:semiparam}

Under Assumption \ref{Ass:FrailtyForIdent}, we propose to use the following proportional hazard specification  for the hazard functions of the six processes 
\begin{align}
\begin{split}
\label{Eq:IDMmodel}
\lambda_{01}(t|a,\bX,\gamma) & = \gamma_a \lambda_{01}^{0}(t|a)\exp(\bX^T\bbeta_{01}^a) \,\,\, , t>0\\
\lambda_{02}(t|a,\bX,\gamma) & = \gamma_a \lambda_{02}^{0}(t|a)\exp(\bX^T\bbeta_{02}^a) \,\,\, , t>0\\
\lambda_{12}(t|t_1,a,\bX,\gamma) &= \gamma_a \lambda_{12}^{0}(t|a)\exp(\bX^T\bbeta_{12}^a) \,\,\, , t>t_1>0 \, ,
\end{split}
\end{align}
where $\lambda_{jk}^{0}(t|a), jk=01, 02,12, a=0,1$ are unspecified baseline hazard functions.
Define also the  cumulative baseline hazard functions, $\Lambda_{jk}^{0}(t|a)=\int_0^t \lambda_{jk}^{0}(u|a)du$, and let $\bbeta = (\bbeta_{01}^{0T},\bbeta_{02}^{0T},\bbeta_{12}^{0T},\bbeta_{01}^{1T},\bbeta_{02}^{1T},\bbeta_{12}^{1T})^T$, and 
$\bLambda_0 = (\Lambda_{01}^{0}(t|a),\Lambda_{02}^{0}(t|a),\Lambda_{12}^{0}(t|a))$ for $a=0,1$. Finally, let $\bpsi=(\bLambda_0,\bbeta, \btheta)$  be the collection of parameters to be estimated from the data.   Assuming the data was prospectively collected with participants being disease-free at study starting time, it can be shown  that the likelihood function \citep{xu2010statistical} is proportional to 
\begin{align}
\label{Eq:Lik}
 \mathcal{L}(\bpsi) = &  \prod_{i=1}^n\Bigg\{ [\lambda_{01}^{0}(\widetilde{T}_{i1}|A_i)]^{\delta_{i1}} [\lambda_{02}^{0}(\widetilde{T}_{i1}|A_i)]^{(1-\delta_{i1})\delta_{i2}} [\lambda_{12}^{0,A_i}(\widetilde{T}_{i2}|A_i)]^{\delta_{i1}\delta_{i2}}\\
 \exp&\bigg[\delta_{i1}\bX_i^T \bbeta^{A_i}_{01} + (1-\delta_{i1})\delta_{i2}\bX_i^T \bbeta^{A_i}_{02} + \delta_{i1}\delta_{i2} \bX_i^T \bbeta^{A_i}_{12}\bigg](-1)^{\delta_{i1}+\delta_{i2}} \phi_{A_i}^{(\delta_{i1} + \delta_{i2})}(s_i)\Bigg\}, \nonumber 
\end{align}
where
$\phi_a(s) = E \left\{ \exp\{ -s \gamma_a \} \right\}$ is the Laplace transform of $\gamma_a$, $a=0,1$, and $\phi_a^{(q)}$ is the $q$th derivative of $\phi_a(s)$ with respect to $s$, $q=0,1,2$, and 
$$
s_i = \sum_{j=1}^2 \exp(\bX_i^T \bbeta_{0j}^{A_i})  \Lambda_{0j}^{0}(\widetilde{T}_{i1}|A_i)  +
\delta_{i1}\exp(\bX_i^T \bbeta_{12}^{A_i}) \big[ \Lambda_{12}^{0}(\widetilde{T}_{i2}|A_i) - \Lambda_{12}^{0}(\widetilde{T}_{i1}|A_i)\big].
$$  
For the likelihood derivation see SM3 of \cite{gorfine2020marginalized}.
We propose to utilize an EM algorithm for maximizing \eqref{Eq:Lik}; the details are given in Section C.3 of the SM.  The algorithm is implemented by our \textbf{R} package \texttt{CausalSemiComp}. Importantly, the estimation phase ignores the unidentifiable component $Cor(\gamma_0,\gamma_1)$ while estimating model \eqref{Eq:IDMmodel} parameters.  Assumptions on the joint distribution of $(\gamma_0,\gamma_1)$ are used to map the estimator $\hat{\bpsi}$ into estimates of causal effects.   Upon estimating $\hat{\bpsi}$, the quantities in the identifying formulas from Proposition \ref{Prop:IdentFrail} can be calculated by numerical integration or Monte Carlo simulations, as we do below in Section \ref{SubSec:NumerSemiParam}.


\section{Simulations}
\label{Sec:Sims}

To assess the finite-sample properties of the proposed  estimators, we conducted simulations under various scenarios with 1,000 simulation iterations per scenario. Sample size was $n=2,000$ and different censoring rates were considered.  The DGM initially followed \eqref{Eq:IDMmodel} with Weibull baseline hazards, two covariates, and  Gamma frailty as described in Example 3 of Section \ref{SubSec:IdentFrail} with $\theta_0=\theta_1$ for simplicity.  Treatment was randomized with $\Pr(A=1)=0.5$.  
Further details on the different DGMs, analyses and results are presented in Section D of the SM. 

\subsection{Non-parametric estimation}
\label{SubSec:NumerNonParam}

We considered the scenarios previously used in Figure \ref{Fig:ACTbounds}.  Assumption \ref{Ass:OrdPersev} was imposed by re-simulating those initially  at the $dh$ stratum.     For the smoothed KM estimator, we used the default choices of the \texttt{prodlim} package \citep{prodlim}. Standard errors were estimated by  bootstrap (sampling at the unit level) with 100 repetitions. 

 The estimators  of  $\eta_{A=a}$, $S_{1|A=a}(10)$ and $S_{1|A=a}(30)$ (Table \ref{Tab:MainResNP}) showed  negligible empirical bias, the standard errors were well estimated  and the 95\% Wald-type confidence intervals had satisfactory empirical coverage rate. Under moderate censoring, an expected increase in the standard errors was observed. Section D.3 of the SM presents results for Scenario (III) and  for additional parameters. 
\begin{table}[ht!]
	\caption{\footnotesize Performance of the proposed non-parametric estimators. Censoring rates  considered  were low (10\%, C-L) and moderate (30\%, C-M).  True: True parameter values; Mean.EST: mean estimate; EMP.SD: empirical standard deviation of the estimates; EST.SE: mean estimated standard error; CP95\%: empirical coverage rate of 95\% confidence interval.\label{Tab:MainResNP}}
\begin{center}
{\footnotesize
\begin{tabular}{lcccccccccccc}
  \hline
 & 	\multicolumn{2}{c}{$\eta_{A=0}$} & 	\multicolumn{2}{c}{$\eta_{A=1}$} & 	\multicolumn{2}{c}{$S_{1|A=0}(10)$} & 	\multicolumn{2}{c}{$S_{1|A=0}(30)$} & 
 \multicolumn{2}{c}{$S_{1|A=1}(10)$} & 	\multicolumn{2}{c}{$S_{1|A=1}(30)$}\\[0.5em] 
 & C-L & C-M &   C-L & C-M &  C-L & C-M & 
    C-L & C-M &  C-L & C-M  &  C-L & C-M \\
 \hline
   &&&&&&\\[-1.5em]
   \multicolumn{3}{l}{Scenario  (I)}&&&&\\[0.25em] 
True &  \multicolumn{2}{c}{0.491} &  \multicolumn{2}{c}{0.848} &  \multicolumn{2}{c}{0.748} &  \multicolumn{2}{c}{0.523} &  \multicolumn{2}{c}{0.481} &  \multicolumn{2}{c}{0.158} \\[0.25em] 
Mean.EST & 0.491 & 0.492  &  0.845 & 0.844 & 0.749 & 0.748 &  0.524 &0.523 & 0.479 & 0.476 &  0.159 &  0.159 \\ 
EMP.SD & 0.016 & 0.020 & 0.011 & 0.012 &  0.014 & 0.016 & 0.016 & 0.020 & 0.017 & 0.019 & 0.012  & 0.012 \\
EST.SE & 0.017 & 0.019 &  0.011 & 0.012 &  0.014 & 0.017 &  0.017 & 0.019 &  0.017 & 0.019 &  0.011 & 0.012 \\
CP95\% & 0.946 &0.944 & 0.951 & 0.948 & 0.950 & 0.949& 0.945 & 0.940 & 0.953 & 0.942 & 0.944 & 0.951 \\ [0.25em]  
\multicolumn{3}{l}{Scenario  (II)}&&&&\\[0.25em] 
True &  \multicolumn{2}{c}{0.144} &  \multicolumn{2}{c}{0.431} &  \multicolumn{2}{c}{0.968} &  \multicolumn{2}{c}{0.867} &  \multicolumn{2}{c}{0.604} &  \multicolumn{2}{c}{0.158} \\[0.25em] 
Mean.EST  & 0.146  & 0.145 & 0.429 & 0.426 &  0.968 & 0.967  & 0.865 & 0.866 &  0.904 & 0.903  & 0.601 & 0.599     \\ 
EMP.SD & 0.012 &  0.015 & 0.016 &   0.018  & 0.006 &   0.007    & 0.012 &   0.014   &  0.010 &    0.012  & 0.016  &   0.019  \\  
EST.SE & 0.012 &   0.015 & 0.016 &   0.019  & 0.006 &   0.007    & 0.012 &   0.015   & 0.010  &   0.013   & 0.016  &   0.019  \\  
CP95\% & 0.945 &  0.946 & 0.938 &   0.937  & 0.939 &    0.942  & 0.943 & 0.949    & 0.953 & 0.956      & 0.945  & 0.941    \\ 
\end{tabular}}
\end{center}
\end{table}

\subsection{Semi-parametric estimation}
\label{SubSec:NumerSemiParam}

 In this simulation study, the data was generated under under model \eqref{Eq:IDMmodel} and was not restricted to follow Assumption \ref{Ass:OrdPersev}.  Frailties were simulated according to Example 3,   with $\rho=0.5$ and equal  variances $\theta=2/3,1,2$, corresponding to Kendall's $\tau=1/4, 1/3, 1/2$ between $T_1(a)$ and $T_2(a)$. For each simulated dataset, we  fitted  two illness-death frailty models for $A=0,1$ using the EM algorithm (Section C.3 of the SM). Frailty variances were estimated separately and then combined (see Section D.2.2 of the SM).  RMST causal effects were estimated by a Monte Carlo integration together with the estimated baseline hazards $\widehat{\bLambda}$, coefficients $\widehat{\bbeta}$ and $\widehat{\theta}$.  This Monte Carlo procedure is described in length in Section D.2 of the SM.  Standard errors   were estimated using  bootstrap with 100 repetitions.

 Generally, the empirical bias of the estimated mean and median RMST  effects was relatively negligible, standard errors were well estimated, and coverage provabilities were satisfactory (Table \ref{Tab:MainResSP}).   The regression coefficients, baseline hazard functions and frailty variance were all also well estimated  (Section D.3 of the SM).  We repeated the simulation studies for $\rho=0,1$ (Section D.3 of the SM); the results did not  change qualitatively. 
\begin{table}
		\caption{\footnotesize Performance of the semi-parametric estimators under Scenario (IV) with $\rho=0.5$,  under  low (5\%, C-L) and moderate (25\%, C-M) censoring rates for $T_2$. Censoring rates for $T_1$ were 35-40\%. The presented effects are \eqref{Eq:estimand:MeanT1ad}, \eqref{Eq:estimand:MeanT2ad} and \eqref{Eq:estimand:MeanT2nd} (ATE) and their median versions (MTE).  True: True parameter values; Mean.EST: mean estimate; EMP.SD: empirical standard deviation of the estimates; EST.SE: mean estimated standard error; CP95\%: empirical coverage rate of 95\% confidence interval.\label{Tab:MainResSP}}
	\centering
	{\footnotesize
	\begin{tabular}{lcccccccccccc}
		\hline
&		\multicolumn{4}{c}{$T_1|ad$} & 		\multicolumn{4}{c}{$T_2|ad$} & 		\multicolumn{4}{c}{$T_2|nd$}\\
&		\multicolumn{2}{c}{ATE} & 	\multicolumn{2}{c}{MTE} &	
		\multicolumn{2}{c}{ATE} & 	\multicolumn{2}{c}{MTE} &
						\multicolumn{2}{c}{ATE} & 	\multicolumn{2}{c}{MTE} \\
 & C-L & C-M &   C-L & C-M &  C-L & C-M & 
C-L & C-M &  C-L & C-M  &  C-L & C-M \\
\hline
$\theta=2/3$ &&&&&&\\
~~True & \multicolumn{2}{c}{-1.16} & \multicolumn{2}{c}{-0.67} & \multicolumn{2}{c}{-3.95} & \multicolumn{2}{c}{-7.17} & \multicolumn{2}{c}{-2.18} & \multicolumn{2}{c}{-1.44} \\ 
~~Mean.EST &  -1.20 & -1.21 & -0.68 & -0.68 & -3.94 & -3.95 & -6.96 & -6.93 & -2.18 & -2.18 & -1.45 & -1.45 \\ 
~~EMP.SD & 0.19 & 0.19 & 0.18 & 0.18 & 0.39 & 0.42 & 0.65 & 0.72 & 0.27 & 0.30 & 0.23 & 0.25 \\ 
~~EST.SE  & 0.18 & 0.19 & 0.18 & 0.19 & 0.40 & 0.43 & 0.65 & 0.72 & 0.27 & 0.29 & 0.24 & 0.25 \\ 
~~CP95\% & 0.94 & 0.94 & 0.94 & 0.95 & 0.95 & 0.95 & 0.96 & 0.93 & 0.95 & 0.93 & 0.95 & 0.94 \\ 
$\theta=1$ &&&&&&\\
~~True & \multicolumn{2}{c}{-1.15} & \multicolumn{2}{c}{-0.71} & \multicolumn{2}{c}{-3.65} & \multicolumn{2}{c}{-6.83} & \multicolumn{2}{c}{-2.42} & \multicolumn{2}{c}{-1.58} \\ 
~~Mean.EST &   -1.20 & -1.20 & -0.73 & -0.73 & -3.66 & -3.68 & -6.72 & -6.69 & -2.42 & -2.44 & -1.59 & -1.61 \\ 
~~EMP.SD &  0.20 & 0.22 & 0.19 & 0.20 & 0.42 & 0.45 & 0.58 & 0.64 & 0.32 & 0.35 & 0.26 & 0.27 \\ 
~~EST.SE & 0.20 & 0.22 & 0.20 & 0.21 & 0.41 & 0.45 & 0.61 & 0.67 & 0.34 & 0.36 & 0.27 & 0.28 \\ 
~~CP95\% & 0.96 & 0.93 & 0.96 & 0.95 & 0.93 & 0.95 & 0.98 & 0.97 & 0.96 & 0.95 & 0.96 & 0.95 \\ 
$\theta=2$ &&&&&&\\
~~True & \multicolumn{2}{c}{-1.01} & \multicolumn{2}{c}{-0.78} & \multicolumn{2}{c}{-3.09} & \multicolumn{2}{c}{-5.60} & \multicolumn{2}{c}{-2.23} & \multicolumn{2}{c}{-2.71} \\ 
~~Mean.EST & -1.02 & -1.04 & -0.78 & -0.80 & -3.07 & -3.06 & -5.59 & -5.55 & -2.31 & -2.30 & -2.78 & -2.74 \\ 
~~EMP.SD & 0.25 & 0.26 & 0.25 & 0.25 & 0.42 & 0.47 & 0.67 & 0.71 & 0.48 & 0.51 & 0.51 & 0.54 \\ 
~~EST.SE &  0.25 & 0.27 & 0.25 & 0.26 & 0.44 & 0.46 & 0.69 & 0.74 & 0.48 & 0.54 & 0.54 & 0.58 \\ 
~~CP95\% &  0.95 & 0.96 & 0.94 & 0.95 & 0.95 & 0.94 & 0.95 & 0.95 & 0.95 & 0.96 & 0.96 & 0.96 \\ 
\end{tabular}}
\end{table}

\section{Illustrative data analysis}
\label{Sec:Data}

The Adult Changes in Thought (ACT) Study is an ongoing prospective cohort study focused on dementia in the elderly. Starting from 1994, Alzheimer's-free participants of age 65 and older from the Seattle metropolitan area  have been recruited. More details can be found  elsewhere \citep{kukull2002dementia,nevo2020modeling}. Here, $A=1$ means having at least one APOE $\epsilon4$ allele. Excluding post-intervention variables, our analyses also include the binary variables gender and race (white/non-white). This leaves us with 4,453 participants, of which 1,783 (40\%) were censored prior to either event, 211 (5\%) were diagnosed with AD and then censored, 1,635 (37\%) died without an AD diagnosis, and 824 (19\%) were diagnosed with AD and died during follow-up.

\begin{figure}[ht!]
	\centering
	\includegraphics[scale = 0.275]{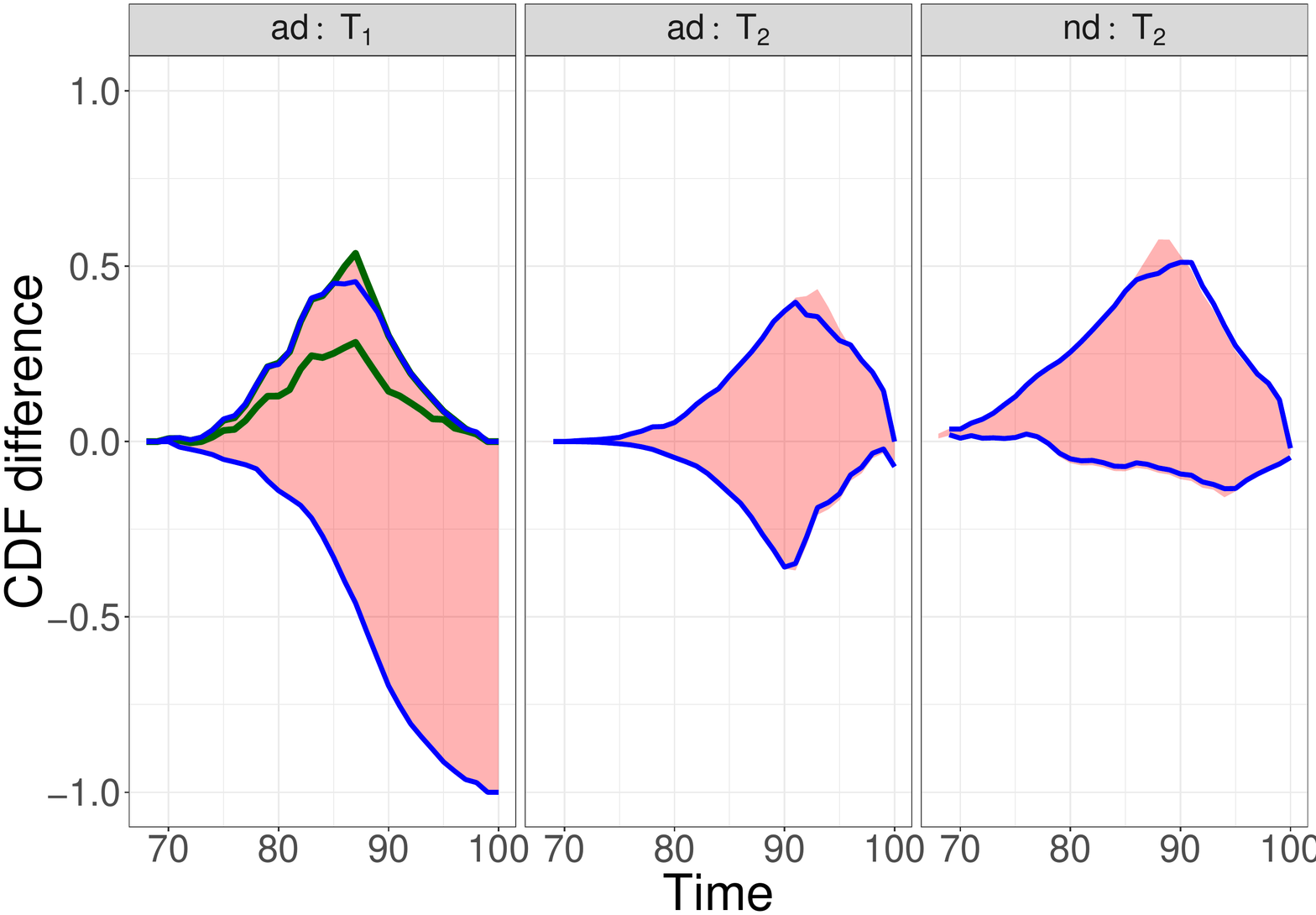}
	\caption{\footnotesize Bounds for principal effects of APOE on AD and death based on the ACT data. In pink shade the simple bounds, in blue bounds adjusted for gender and in green bounds based on Assumption \ref{Ass:Rank}.}
	\label{Fig:ACTbounds}
\end{figure}

Figure \ref{Fig:ACTbounds} presents the estimated bounds for  causal effects \eqref{Eq:estimand:ProbT2ad}--\eqref{Eq:estimand:ProbT1ad} under Assumption \ref{Ass:OrdPersev} (in pink shade), utilizing gender as an additional covariate (in blue) and under the additional Assumption \ref{Ass:Rank} (in green, for \eqref{Eq:estimand:ProbT1ad} only). Both Assumptions \ref{Ass:OrdPersev} and  \ref{Ass:Rank} seem reasonably plausible in this application, because it is well-established that APOE is a strong predictor of AD.  The $ad$ proportion $\pi_{ad}$ was estimated to be 35\% (CI95:  32\%, 37\%). The bounds $[\widetilde{\mathcal{L}}_{1,ad}(t), \mathcal{U}^{Z}_{2,ad}(t)]$ (bottom green and top blue) demonstrate that APOE  induces earlier AD onset within the $ad$ stratum. The bounds do not allow for definite conclusions regarding the effect of APOE on death, and the inclusion of the variable gender had only minor effect. Focusing on the $nd$ stratum, its proportion was estimated to be 53\% (CI95:  49\%, 58\%), meaning the two stratum of interest together comprise the vast majority of the population. Whether APOE had a positive or negative effect on survival in the $nd$ is unclear, although the lower bound was larger than zero for early ages.

In the semi-parametric analysis under Assumption \ref{Ass:FrailtyForIdent} and Example 3, two illness-death models \eqref{Eq:IDMmodel} were fitted using the EM algorithm, and included gender and race as covariates $\bX$ (see SM Table E.14  for the estimated coefficients).  Because participants in the ACT were recruited in varying ages, data was left truncated. We followed  the approximation of \cite{nielsen1992counting}, by replacing all risk indicators $I\{t \le \widetilde{T}_{ij}\}$ with $I\{R_i \le t \le \widetilde{T}_{ij}\}$, where $R_i$ is the age at recruitment.  Estimated Gamma frailty variances were $\hat{\theta}_0=0.047,\hat{\theta}_1=0.027$ and then combined for simplicity of Monte Carlo estimation of casual effects. The estimated frailty variance was  $\hat{\theta}=0.042$ (SE: $0.005$), corresponding to Kendall's $\tau$ of 0.02. Turning to causal effects (Table \ref{Tab:DataRes}), significant long-term  negative effects of APOE on AD onset within the $ad$ stratum were estimated; those  with APOE are expected to receive an AD diagnosis approximately 2 years earlier. Additionally, death at earlier age was expected in both strata under APOE.  For $t^\star=15$ (age 80),  effects were zero for the $nd$ stratum and negative though small for the $ad$. For $t^{\star}=25,35$ (ages 90,100), effects were negative in both strata, but stronger for the $ad$.    The estimated effects did not change substantially as a function of $\rho$, as expected in low frailty variance scenarios (recall Figure \ref{Fig:FrailtyCausalEffects}).

To summarize the findings,  APOE was found to expedite AD development  within the $ad$ stratum.  APOE was shown to have harmful effect on long-term age at death in both strata only under Assumption \ref{Ass:FrailtyForIdent} and the semi-parametric modeling.

\begin{table}[ht!]
\centering
		\caption{\footnotesize Estimated RMST causal effects of APOE on AD and death times within the $ad$ and $nd$ strata. Effects presented are  mean (ATE) and median (MTE) differences. Results presented for $\rho=0.5$. The empirical distribution of gender and race in the data was used for calculating RMST effects.\label{Tab:DataRes}}
		\footnotesize
\begin{tabular}{cccccc}
  \hline
& &		\multicolumn{2}{c}{AD}  	&	\multicolumn{2}{c}{Death} \\
Stratum & $t^\star$& $\widehat{ATE}$ (95\% CI) & $\widehat{MTE}$ (95\% CI)  & $\widehat{ATE}$ (95\% CI)  &  $\widehat{MTE}$ (95\% CI)  \\
  \hline
$ad$ &$15$ &  -0.04 (-0.12, 0.04) & 0.00 (-0.07, -0.52) & -0.61 (-0.70, -0.05) & -1.00 (-1.09, 0.08) \\ 
 & $25$ & -1.28 (-1.35, -1.20) & -2.00 (-2.07, -2.43) & -2.52 (-2.61, -0.48) & -3.00 (-3.08, -0.92) \\ 
& $35$  & -1.82 (-1.89, -1.74) & -2.01 (-2.09, -2.35) & -2.42 (-2.5, -1.67) & -3.00 (-3.07, -1.90) \\ 
$nd$ & $15$ &  &  & -0.13 (-0.22, 0.07) & 0.00 (-0.08, -0.91) \\ 
& $25$ &  &  & -0.56 (-0.65, -1.92) & -1.00 (-1.08, -2.92) \\ 
& $35$ &  &  & -1.75 (-1.84, -1.92) & -1.99 (-2.08, -2.93) \\ 
   \hline
\end{tabular}
\end{table}

\section{Discussion}
\label{Sec:discuss}

It has been increasingly acknowledged that  considering dual outcomes may provide more information about the scientific problem in hand, especially in the presence of death as a competing risk. In time-to-event data analysis, the semi-competing risks framework allows for consideration of dual outcomes. However, while methods have been developed for studying such data, questions of causality were only recently began to be studied \citep{comment2019survivor,xu2020bayesian}.

We proposed new estimands built upon a population stratification approach, and discussed their partial and full identifiability, under variety of assumptions. The utility of the proposed approach is  largest when either the never-diseased or the always-diseased strata comprise many of the population members. As demonstrated in Section \ref{Sec:AssumpPrtIdent}, in these cases the partial identification is sharper for the population-stratified effects.  This paper established an approach to study causal effects for semi-competing risks data, and discussed both the theory it hinges on and its practical use. Ultimately, we believe this will strengthen semi-competing risks data analysis, giving clear guidelines to what causal analyses should be carried out and under which  assumptions causal conclusion can be derived.

\vspace{-0.5cm}
	\bibliographystyle{chicago}
\bibliography{CausalSemiComp}

\end{document}